\documentclass[preprint,12pt, authoryear]{elsarticle}

\usepackage[margin=1in]{geometry}

\usepackage{amsmath}

\usepackage{natbib}

\usepackage{amssymb}

\usepackage{todonotes}
\usepackage{hyperref}

\makeatletter
\def\ps@pprintTitle{%
  \let\@oddhead\@empty
  \let\@evenhead\@empty
  \def\@oddfoot{\reset@font\hfil\thepage\hfil}
  \let\@evenfoot\@oddfoot
}
\makeatother

\begin{document}

\begin{frontmatter}



\title{Accurate calculation of bubble and droplet properties in diffuse-interface two-phase simulations}


\author[]{Pranav J. Nathan}

\affiliation[]{organization={Flow Physics and Computational Science Lab, George W. Woodruff School of Mechanical Engineering, Georgia Institute of Technology},
            city={Atlanta},
            state={GA},
            country={USA}}

\author[]{Suhas S. Jain\corref{cor1}}
\cortext[cor1]{Corresponding author: Suhas S. Jain (suhasjain@gatech.edu)}

\begin{abstract}
In this paper, we address the challenge of accurately calculating droplet/bubble properties (e.g., volume, number) in diffuse-interface two-phase flow simulations. Currently, flood-fill algorithms can truncate a significant portion of the volume of droplets/bubbles contained within the diffuse interface region or artificially merge multiple droplets/bubbles. This error is also dependent on the volume fraction cutoff value, which is typically chosen to be 0.5 arbitrarily, in the flood-fill algorithms. We propose a simple volume-correction approach that incorporates an analytical approximation of the truncated volume to correct for the missing droplet/bubble volumes. This proposed method results in accurately recovering the dispersed phase volumes with minimal volume error over a wide range of volume fraction cutoff values, and hence, can also accurately recover the number of droplets/bubbles. This can be a valuable tool for accurate calculation of drop/bubble size distributions for analysis and for Eulerian-to-Lagrangian conversion of the dispersed phase in multi-scale modeling approaches. 
\end{abstract}


\end{frontmatter}

\section{Introduction}
\label{sec:intro}
Modeling of two-phase flows and the study of the deformation, breakup and coalescence of bubbles and droplets is an important problem relevant to various engineering and environmental processes [\cite{brennen2005fundamentals}]. The atomization of jets and sprays, emulsions, boiling phenomena, and carbon sequestration all involve complex interactions between fluid phases [\cite{gorokhovski2008modeling}; \cite{PAL201141} \cite{boiling} \cite{JIANG20113557}]. In both laminar and turbulent settings, these interactions play a significant role in shaping the dynamics between the dispersed and carrier phases. As such, it is critical to understand how these interfaces evolve, since the exchange of mass, momentum, and energy between the phases occur at the interface. 

In these two-phase flows, it is critical to determine accurate values of the volume, surface area, number of droplets/bubbles, and droplet-size distribution (DSD). In engineering applications, these flow characteristics derive essential design parameters, such as the Sauter mean diameter (SMD) of fuel droplets in a gas turbine combustor. For environmental applications, they can determine the amount of air trapped in breaking waves, the drag experienced by raindrops, and the rate of dissolution of bubbles in an aqueous solution. In addition to these direct applications in engineering and nature, accurate determination of dispersed phase volumes in two-phase flows is also important for conversion from Eulerian-to-Lagrangian phase for Euler-Lagrange simulations of these flows. 

Currently, the flood-fill algorithm represents one of the best ways to calculate the volume of droplets in a two-phase flow. It excels in circumstances where there is a sharp interface between the two phases of the flow as was described in \cite{LU2015123} and \cite{HERRMANN2010745}. However, when applied to a two-phase flow field with a diffuse interface, it struggles to accurately capture the volumes due to the lack of a clear distinction between one phase and another. {Because the volume fraction varies smoothly between the two phases, there is no clear cutoff value. In contrast, sharp-interface methods have a clear boundary that can distinguish between the two phases easily.} Using the flood-fill algorithm and choosing an arbitrary cut-off value of the volume fraction could result in undesirable effects. Choosing a higher value of the cut-off results in significant underprediction of the volume of drops/bubbles, and choosing a lower value could result in artificial merging of separate drops/bubbles. Hence, in this paper, we propose a simple approach for accurate calculation of volumes of drops and bubbles in two-phase flows. We propose an analytical correction for the missing volumes, when the classical flood-fill algorithms are used to detect dispersed phases in diffuse interface approaches, which results in accurate calculation of volumes of drops and bubbles for wide range of cut-off values. With this new approach, a high enough cut-off value of volume fraction can then be used to prevent artificial merging of drops/bubbles while still accurately recovering the volume of drops and bubbles. The proposed approach is equally applicable for both drops and bubbles, but for brevity, the dispersed phase is referred to as droplets in the rest of the paper. 

In Section \ref{sec:pfmethod}, the phase-field method that is used to model two-phase flows is described. In Section \ref{sec:floodfill}, the limitations of the flood-fill algorithm for this application are further elaborated on. In Section \ref{sec:approaches}, the three different approaches for volume calculation are explained, including the proposed approach. Finally, in Section \ref{sec:results}, the three described approaches are applied to sample flow fields and their performance is compared, followed by conclusions in Section \ref{sec:conclusions}. 

\section{Phase-field method}\label{sec:pfmethod}

\begin{figure}[h!]
    \centering
    \includegraphics[width=0.55\linewidth]{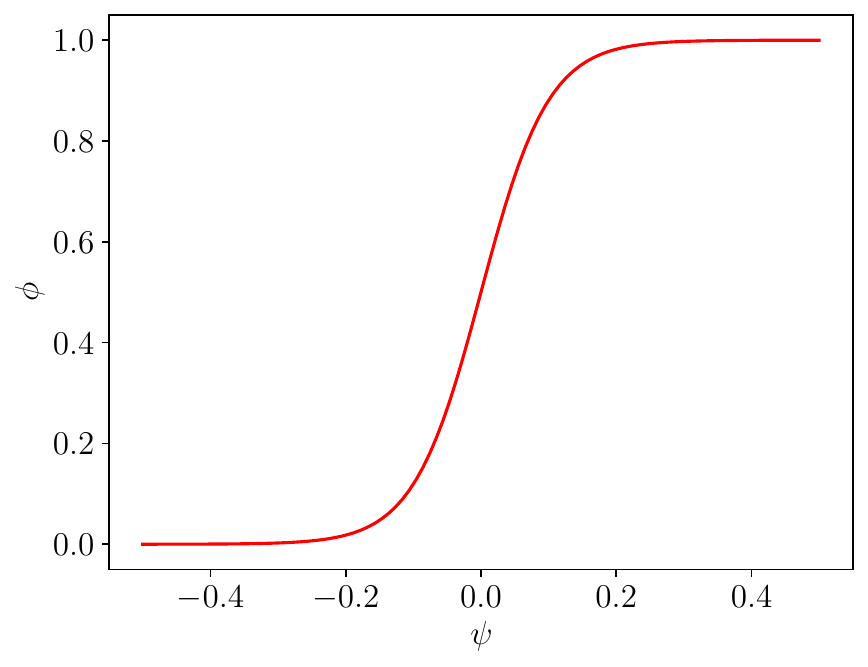}
    \caption{The interface transitions from $\phi$ = 1 to 0 gently following a hyperbolic tangent model.}
    \label{fig:equilibrium}
\end{figure}

A phase-field (PF) method is a diffuse-interface approach for modeling the dynamic topological changes inherent to two-phase flows. PF methods are comparatively less expensive than other interface-capturing methods and are simple to implement while resulting in conservative transport of two phases. 
The challenge with calculating the number of droplets and volume within any general diffuse-interface model is the lack of a numerically sharp interface. Rather, the diffuse-interface method models the interface as approximately a hyperbolic tangent as a function of distance along the interface normal direction (see Figure \ref{fig:equilibrium}) given by
\begin{equation}
\phi \left(\psi\right) = \frac{1}{2} \left[1+\tanh\left(\frac{\psi}{2\epsilon}\right)\right], 
\label{eq:hyperbolic-tangent}
\end{equation}
where $\psi$ is the signed normal distance from the interface. 
This diffuse nature of the interface introduces complexity when post-processing the solution fields due to the lack of clear distinction between the phases, as described in detail in Section \ref{sec:intro}.

In this paper, we utilize the simulations performed using the accurate conservative diffuse-interface/phase-field (ACDI/ACPF) model \citep{JAIN2022111529}. The ACDI model for $\phi$, volume fraction of one of the fluids, can be written as
\begin{equation}
\frac{\partial \phi}{\partial t} + \vec \nabla \cdot {(\phi \vec{u})} = \vec \nabla \cdot \left\{ \Gamma \left\{ \epsilon \nabla \phi - \frac{1}{4} \left[ 1 - \tanh^2 \left( \frac{\psi}{2\epsilon}\right)\right]  \frac{\nabla \psi}{|\nabla \psi|}\right\}\right\},
\label{eq:acdi}
\end{equation}
where $\epsilon$ represents an interface thickness scale and $\Gamma$ represents artificial regularization velocity scale.

This recently developed ACDI method has gained popularity because of its suitability for accurate modeling of complex interface dynamics in drop-laden turbulence \citep{JAIN2022111529, jain2025stationary, PhysRevFluids.8.090501, HWANG2024112972} and a wide range of other multiphysics flow problems \citep{scapin2022mass, brown2023phase, jain_solsurf2023, JAIN2024113277}. It is known to be more accurate than other existing phase-field models because it maintains a sharper interface\textemdash with only one-to-two grid points across the interface\textemdash while being non-dissipative, robust, less expensive, and conservative, without the need for any geometric treatment. Because of these advantages, the ACDI method was chosen as the interface-capturing method for the simulations in this work. 

However, the methods described in this paper apply to all diffuse interface approaches, including phase-field methods and algebraic volume-of-fluid models, especially those that admit equilibrium interface profiles. 
Some examples of the other commonly used diffuse interface approaches that have an equilibrium analytical expression for the interface between two phases are: the Cahn-Hilliard model \citep{10.1063/1.1744102, Magaletti_Picano_Chinappi_Marino_Casciola_2013, CHEN2020109782, KHANWALE2020109674}, Allen-Cahn model \citep{ALLEN19791085}, conservative diffuse-interface model \citep{CHIU2011185} and its compressible variants in \citet{JAIN2020109606, JAIN2023111866} with five-and four-equation models, and conservative Allen-Cahn model \citep{brassel2011modified, HUANG2020109718} and its compressible variant in \citet{HUANG2023112195}. 

\section{Flood-fill algorithm and its limitations}\label{sec:floodfill}

Currently, the best practice for efficiently calculating the volume of an individual drop is using flooding algorithms. The flood-fill algorithm is a recursive procedure used to determine and mark all contiguous regions within a multi-dimensional space that share a specific property, commonly employed in image processing to fill connected areas with a uniform color or pattern \citep{10.1145/965103.807456}. 

\begin{figure}[h!]
    \centering
        (a)
        \includegraphics[width=0.45\textwidth]{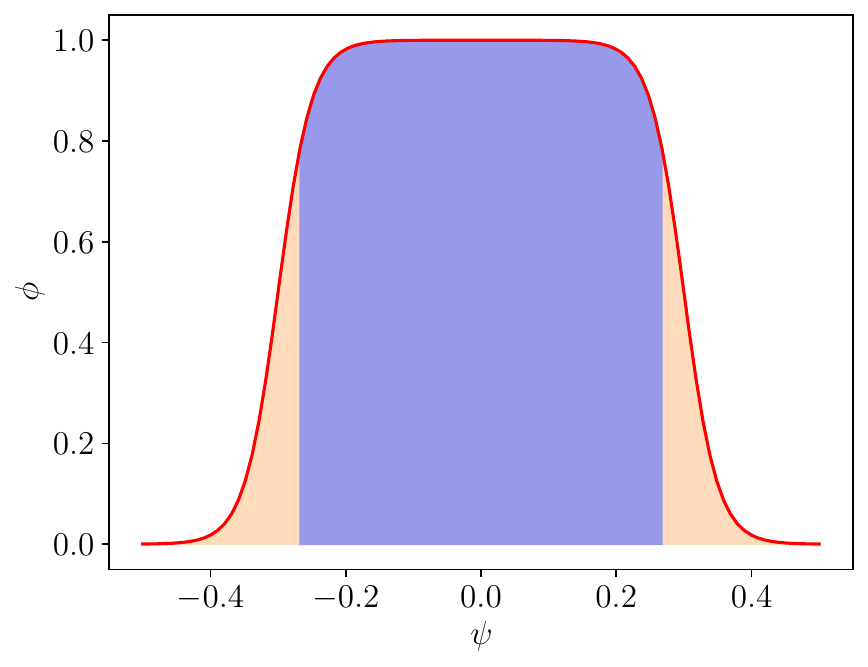}
        (b)
        \includegraphics[width=0.45\textwidth]{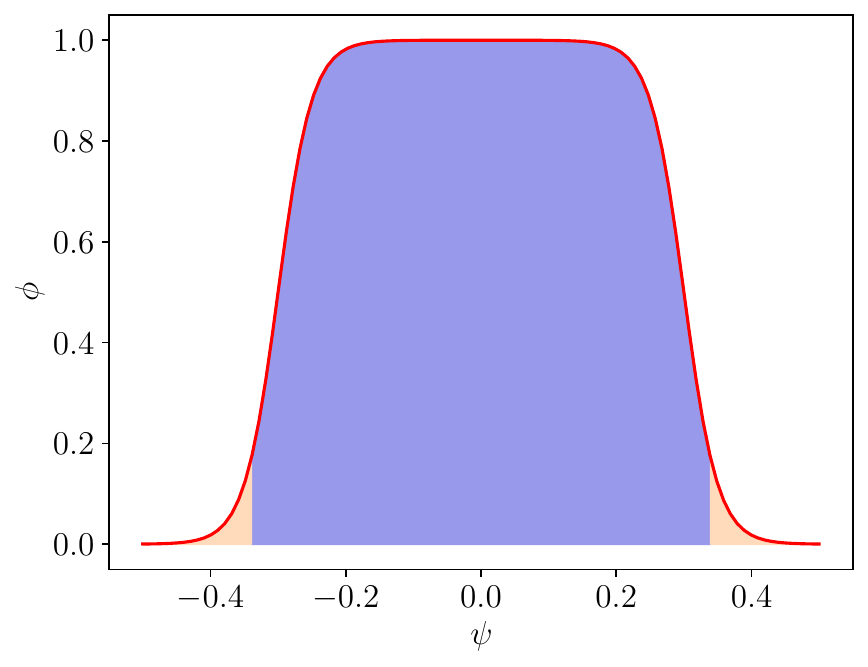}
      { \caption{A schematic showing droplet(s) using volume fraction fields in red. The blue shaded region is the one that will be used in the flood-fill algorithm to compute the droplet volume. (a) Here $\phi_c=0.8$, and the orange region is the region not accounted for in the flood-fill algorithm, which results in an error in the droplet volume. (b) Here $\phi_c=0.2$, is lower, and as a result, reduces the error in the tails. This presents challenges discussed below.}}
        \label{fig:fflimits1}
\end{figure}

The flood-fill algorithm can be applied for volume calculation in two-phase flows by recursively checking all nearby grid points against a certain cutoff value of volume fraction, $\phi_c$.
When $\phi_c$ is large, this approach will result in neglecting a large portion of the volume of the drop, as shown in Figure \ref{fig:fflimits1}(a). A smaller choice of cutoff value will result in a lower volume error [Figure \ref{fig:fflimits1}(b)]. 

\begin{figure}[h!]
    \centering
        \includegraphics[width=0.45\textwidth]{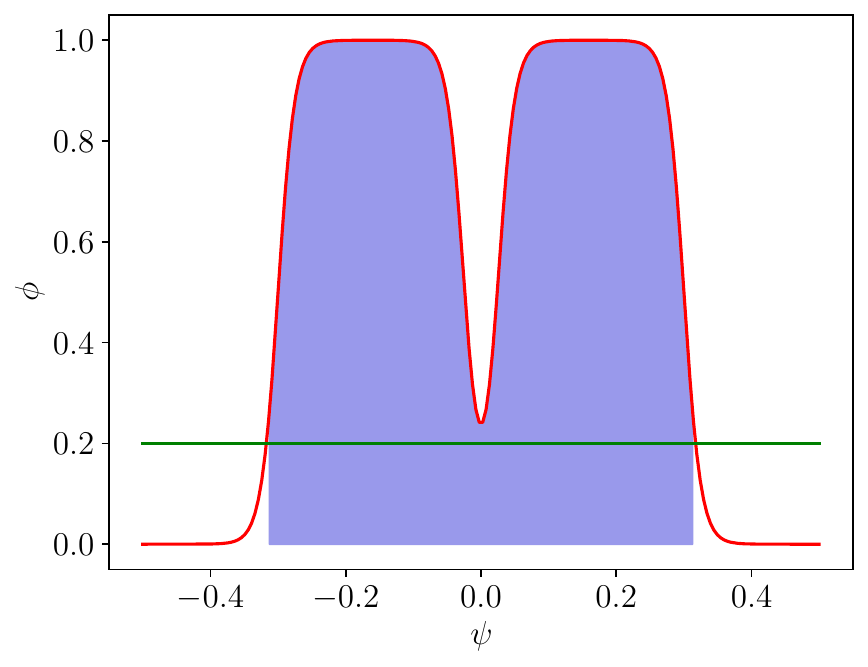}
        \caption{A schematic showing droplet(s) using volume fraction fields in red. The blue shaded region is the one that will be used in the flood-fill algorithm to compute the droplet volume. Here $\phi_c=0.2$, which is too low, and as a result, multiple nearby drops can be artificially merged.}
        \label{fig:fflimits2}
\end{figure}

However, if the cutoff value is too low, the probability of counting two distinct but nearby droplets as a single larger droplet increases considerably, as illustrated in Figure \ref{fig:fflimits2}. A new method is thus needed that addresses this limitation with the flood-fill algorithm.

\section{Volume calculation approaches}\label{sec:approaches}

In this section, we present three approaches, including the proposed approach, to calculate the volume of droplets in two-phase flows.

\subsection{Approach 1 (classical flood fill)}

Approach 1 estimates the volume of droplets by masking the droplets using a cutoff value, $\phi_{c}$. From there, the flood-fill algorithm can be used to calculate the volume of the connected regions of fluid by setting a value of 1 inside the masked region, as illustrated in Figure \ref{fig:approach12}(a), and by integrating this value within each of the masked regions. The fluid volume calculated by this approach is likely to either be over or under-predicted depending on the chosen value of $\phi_{c}$. This is because the shaded area in Figure \ref{fig:approach12}(a) may not be equal to the area under the volume fraction curve.

\subsection{Approach 2}

\begin{figure}[h!]
    \centering
        (a) \includegraphics[width=0.45\textwidth]{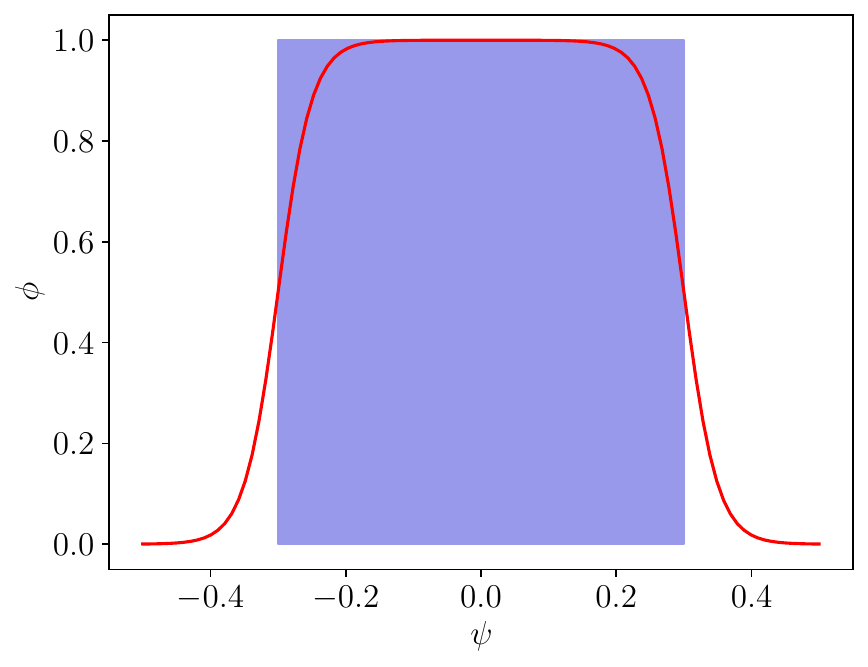}
        (b) \includegraphics[width=0.45\textwidth]{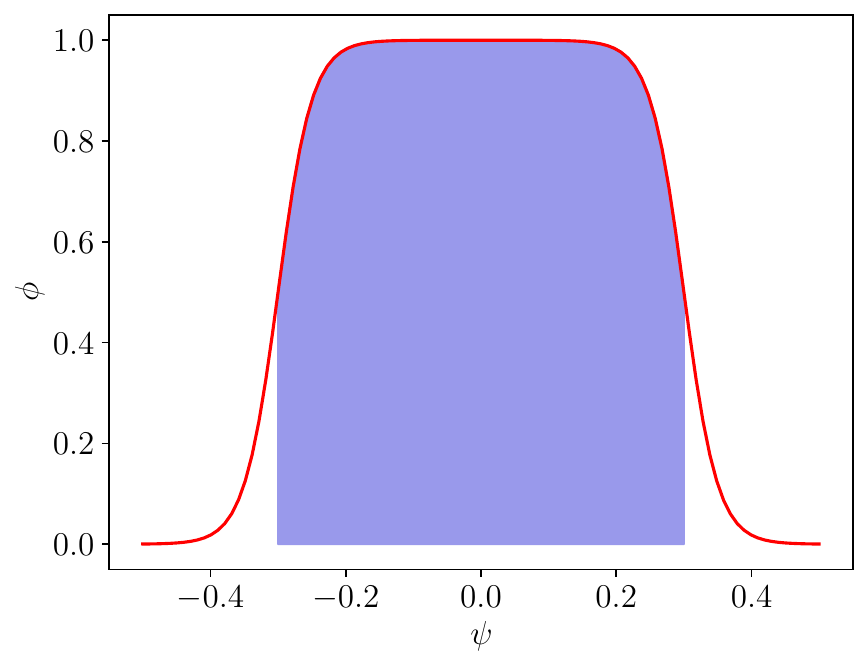}
        \caption{A schematic showing droplet(s) using volume fraction fields in red and the calculated droplet fluid volume in blue. (a) Approach 1 is schematically shown, where all positions within the cutoff $\phi$ range of 0.5 are masked and valued at 1, and then summed up to calculate the volume of the droplet. (b) Approach 2 is schematically shown, where all positions within the cutoff $\phi$ range of 0.5 are masked, and the volume fraction values inside the region is summed up to calculate the volume of the droplet.}
        \label{fig:approach12}
\end{figure}

Approach 2 uses the same masking and flood-fill algorithm, as in approach 1, to calculate the volume of the droplets. But instead of evaluating a value of 1 inside the masked region, as in approach 1, it sums up the volume fraction values inside the masked region [see Figure \ref{fig:approach12}(b)]. This approach more accurately represents the volume of the drop, preventing the overestimation of the volume contributions at the surface of the droplet. 

Approach 2 will no longer overestimate the total fluid volume, but will diverge from the true value as you increase the $\phi_c$ value. This is due to the volume contributions at the surface of the droplets being cut off during the masking procedure, as shown in the schematic in Figure \ref{fig:approach12}(b). 

\subsection{Proposed approach}

\begin{figure}[h!]
    \centering
    \includegraphics[width=0.55\linewidth]{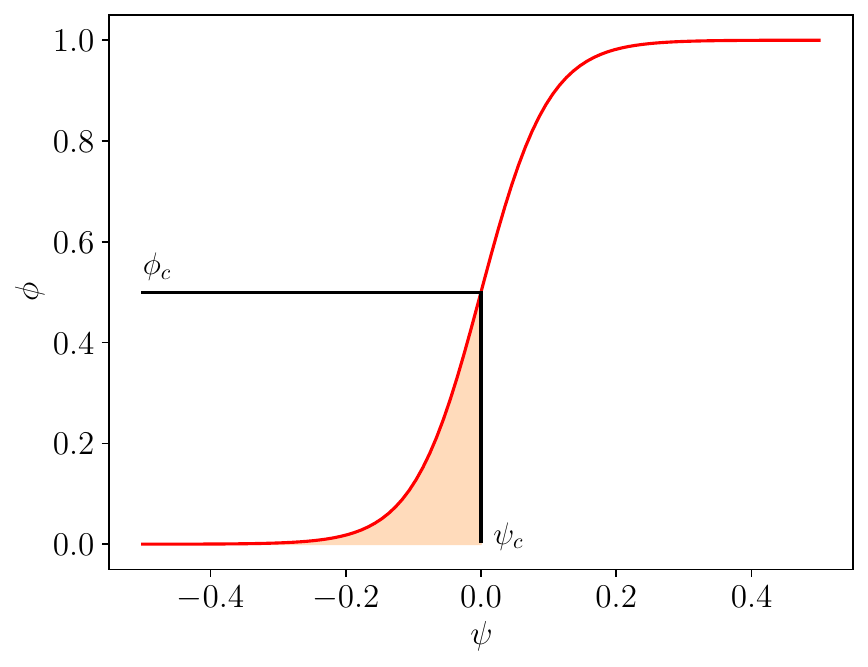}
    \caption{Zoomed in view of the interface transition from $\phi$ = 1 to 0, following a hyperbolic tangent model. The shaded area represents the integrated tail of the droplet, $I_T$.}
    \label{fig:tailint}
\end{figure}

The proposed approach builds off of approach 2 by taking the volume calculated within the $\phi_c $ cutoff value and adding an approximation of the volume contained within the truncated diffuse interface region. The first step of this process is to calculate an approximation of the surface area of an individual droplet. {An ellipsoid droplet approximation is made for this. To calculate the surface area of a droplet, the ellipsoid axes are determined using the inertia eigenvalues \citep{lajevardi2011geoobjects}. From there, the surface area of the ellipsoid can be approximated using the Thomsen-Michon approximation as} 
\begin{equation}
SA_{drop,2} \;\approx\; 4\pi
\!\left(
    \frac{(ab)^{1.6075} + (ac)^{1.6075} + (bc)^{1.6075}}{3}
\right)^{\!1/1.6075},
\label{eq:ellipsoidSA}
\end{equation} 
{where $a$, $b$, and $c$ are the semi-major axes of the ellipsoid. One can also approximate the surface area of a droplet using spherical droplet approximation, which is described in Appendix A.}
The next step is to calculate the volume contained in the truncated diffuse interface region, which can be analytically approximated by first integrating the area under the volume fraction curve from $-\infty$ to $\psi(\phi_c)$ as illustrated in Figure \ref{fig:tailint} to calculate the interface thickness $I_T$ as 
\begin{equation}
I_T = \int_{-\infty}^{\psi\left(\phi_{c}\right)} \frac{1}{2}\left[1 + \tanh\left(\frac{\psi}{2\epsilon}\right)\right]d\psi = \epsilon\ln\left\{\cosh\left[\frac{\ln\left(\frac{-\phi_{c}}{\phi_{c} - 1}\right)}{2}\right]\right\} + \frac{1}{2}\epsilon\ln\left(\frac{-\phi_{c}}{\phi_{c} - 1}\right) + \epsilon\ln\left(2\right).
\label{eq:integral}
\end{equation} 
Here $\psi(\phi_c)$ represents the location that corresponds to the cut-off value of $\phi$ and can be calculated by inverting Eq. \eqref{eq:hyperbolic-tangent} to find $\psi$ \citep{JAIN2022111529} as
\begin{equation}
\psi\left(\phi \right) = \epsilon\ln\left({\frac{\phi + \varepsilon}{1 - \phi + \varepsilon}}\right),
\label{eq:psi}
\end{equation}
where $\varepsilon$ is a small number that is added, to both the numerator and the denominator, to avoid $\psi$ going to $-\infty$ or $\infty$ when $\phi$ goes to 0 or 1, respectively (a value of $\varepsilon = 10^{-100}$ is used in this work).

\begin{figure}[h!]
    \centering
    (a)
    \includegraphics[width=0.35\linewidth]{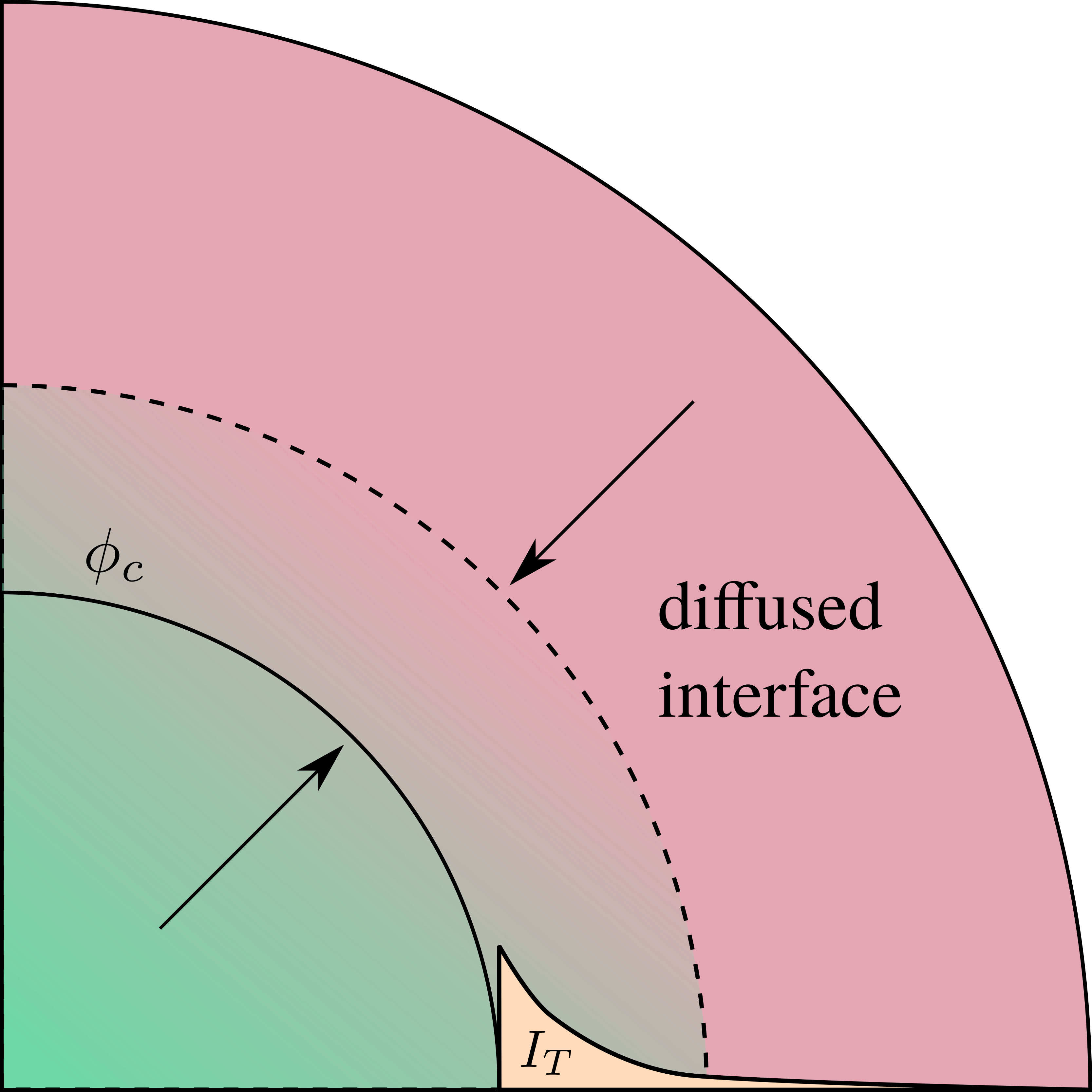}
    (b)
    \includegraphics[width=0.45\linewidth]{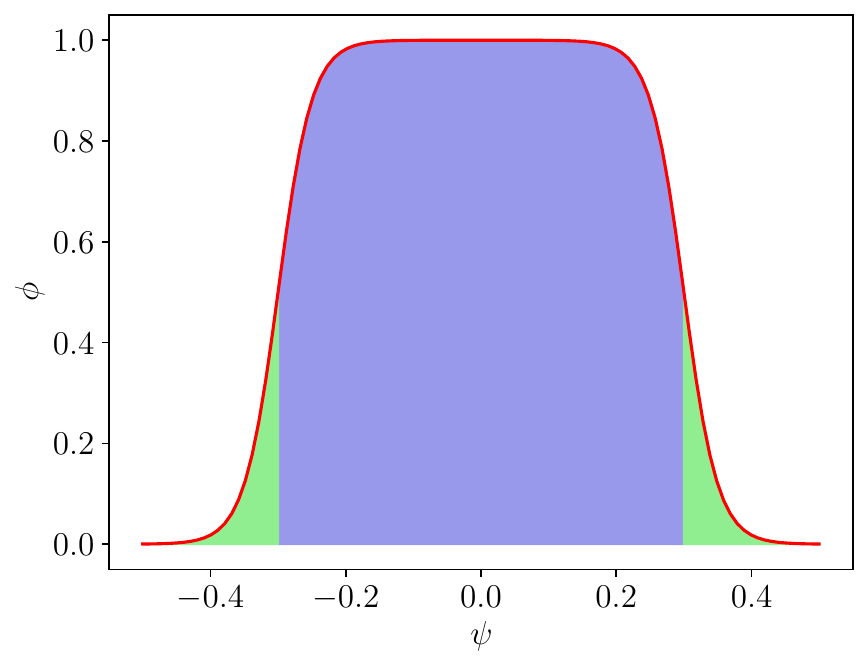}
    \caption{A schematic showing the proposed approach, which includes a correction to the volume by adding the volume contained within the tails. (a) An illustration of the proposed approach in 2D, showing the multiplication of the droplet surface area with the interface thickness, $I_T$, to calculate the truncated volume. (b) A 1D representation of the droplet(s) using volume fraction fields in red and the calculated fluid volume in blue and the correction in green.}
    \label{fig:proposed-approach}
\end{figure}

Finally, the interface thickness value, $I_T$, calculated using Eq. \eqref{fig:tailint} is multiplied by the surface area of the drop [Eq. \eqref{eq:dropletSA}] to calculate the truncated volume that is contained in the diffuse interface region [as illustrated in Figure \ref{fig:proposed-approach}(b)] and then added to $V_{droplet}$ [as illustrated in Figure \ref{fig:proposed-approach}(a)] as 
\begin{equation}
    V_{drop,3} = V_{drop,2} + V_{truncated} = V_{drop,2} + I_T * SA_{drop,2}.
    \label{eq:V_drop3}
\end{equation}
This process is then repeated for each of the droplets in the flow field. {A 2D version of the proposed method for application in 2D simulations is presented in Appendix B.}

\section{Results}\label{sec:results}

We will explore two different simulation setups to evaluate the accuracy of the proposed method and compare it with the classical approaches. In the first setup, we will consider a single stationary droplet case in Section \ref{sec:results1}. Following that, we will consider a turbulent setup with multiple droplets in Section \ref{sec:results2}.

\subsection{Single-droplet case \label{sec:results1}}

\begin{figure}[h!]
    \centering
    \includegraphics[width=0.5\linewidth]{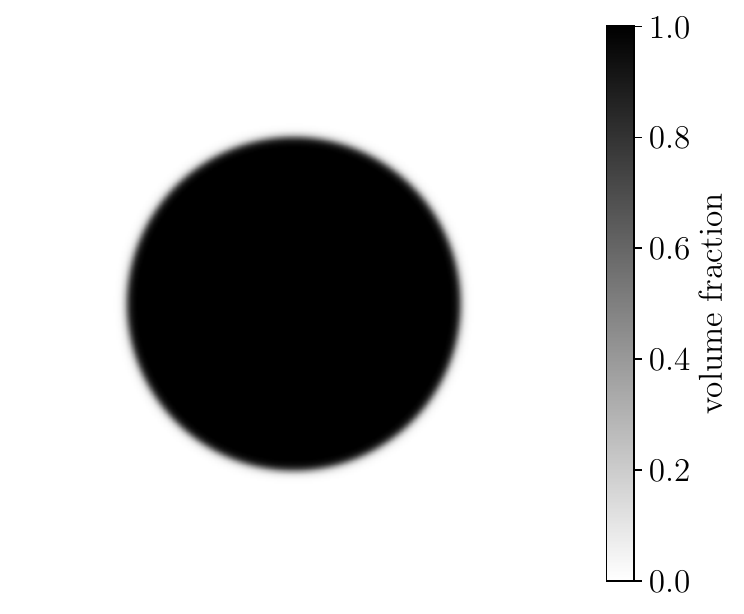}
    \caption{This visualization shows a cross-section of a diffuse-interface spherical droplet located at the center of the domain.}
    \label{fig:singledrop}
\end{figure}

In this section, a single three dimensional stationary droplet was initialized at the center of a cubic domain with side length $L$ as shown in Figure \ref{fig:singledrop}. It contains a diffuse interface between the droplet phase and the carrier phase. The droplet was initialized with a diameter of $0.6 L$, and the value of $\epsilon$ was chosen to be $0.51\Delta$, where $\Delta= L/250$ is the grid size.

\begin{figure}[h!]
    \centering
    \includegraphics[width=\linewidth]{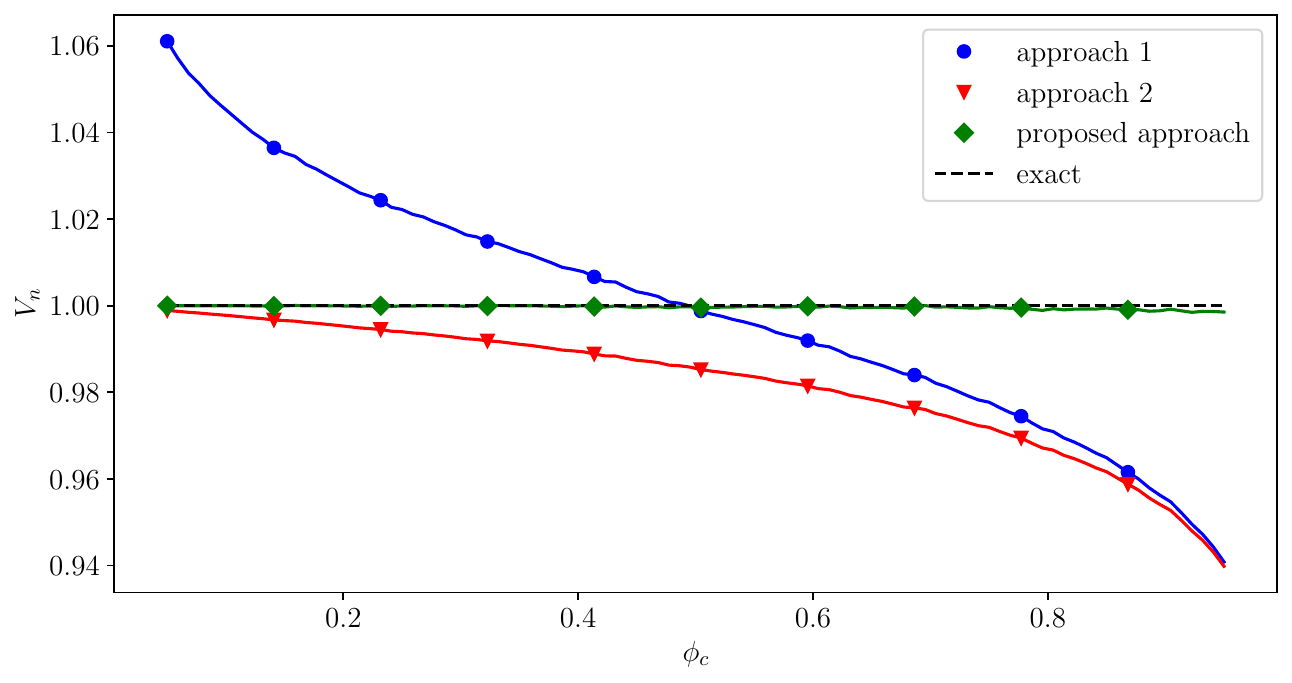}
    \caption{The calculated normalized droplet volume, $V_n$, as a function of volume fraction cut-off values for the single droplet. The proposed approach outperforms the other approaches over all values of $\phi_c$.}
    \label{fig:singapproaches}
\end{figure}

The three approaches described in Section \ref{sec:approaches} (which includes the proposed approach) were applied to the stationary droplet case to assess their performance. {Performance was evaluated by comparing the total sum of the droplet volumes from each of the approaches against the exact total dispersed phase volume in the flow. This exact volume, $V_{exact}$, is computed as}
\begin{equation}
V_{exact} = \sum_i \phi_i.
\label{eq:V_true}
\end{equation}
The plot in Figure \ref{fig:singapproaches} shows the normalized volume, $V_n$, calculated using the three methods for a range of $\phi_c$ values. $V_n$ is the calculated volume normalized by the exact total droplet volume, defined as
\begin{equation}
V_{n} = \frac{V_{drop,k}}{V_{exact}},
\label{eq:V_n}
\end{equation}
where $V_{drop,k}$ is the calculated volume of the droplet(s) using approach $k$, $\phi_i$ is the value of the dispersed phase volume fraction at a particular grid point $i$. A value $V_n$ of 1 represents an accurate volume calculation approach that results in the exact value of the dispersed phase volume in the simulation and is the target value for all approaches. As seen in Figure \ref{fig:singapproaches}, approach 1 results in overprediction of droplet volume for lower values of $\phi_c$ and underprediction for higher values; approach 2 results in underprediction for all values of $\phi_c$; and the proposed approach accurately calculates the droplet volume for all values of $\phi_c$, except a slight underprediction for very large values of $\phi_c$. {This is because when $\phi_c$ is close to one, the correction approximation is performing the majority of the calculation, amplifying the impact of numerical errors.} The effect of the grid resolution on the accuracy of the proposed approach is presented in Appendix C{, and the effect of the interface thickness, $\epsilon$, on the accuracy is presented in Appendix D}.

\subsection{Multiple droplets in turbulence \label{sec:results2}}

\begin{figure}[h!]
    \centering
    \includegraphics[width=0.8\linewidth]{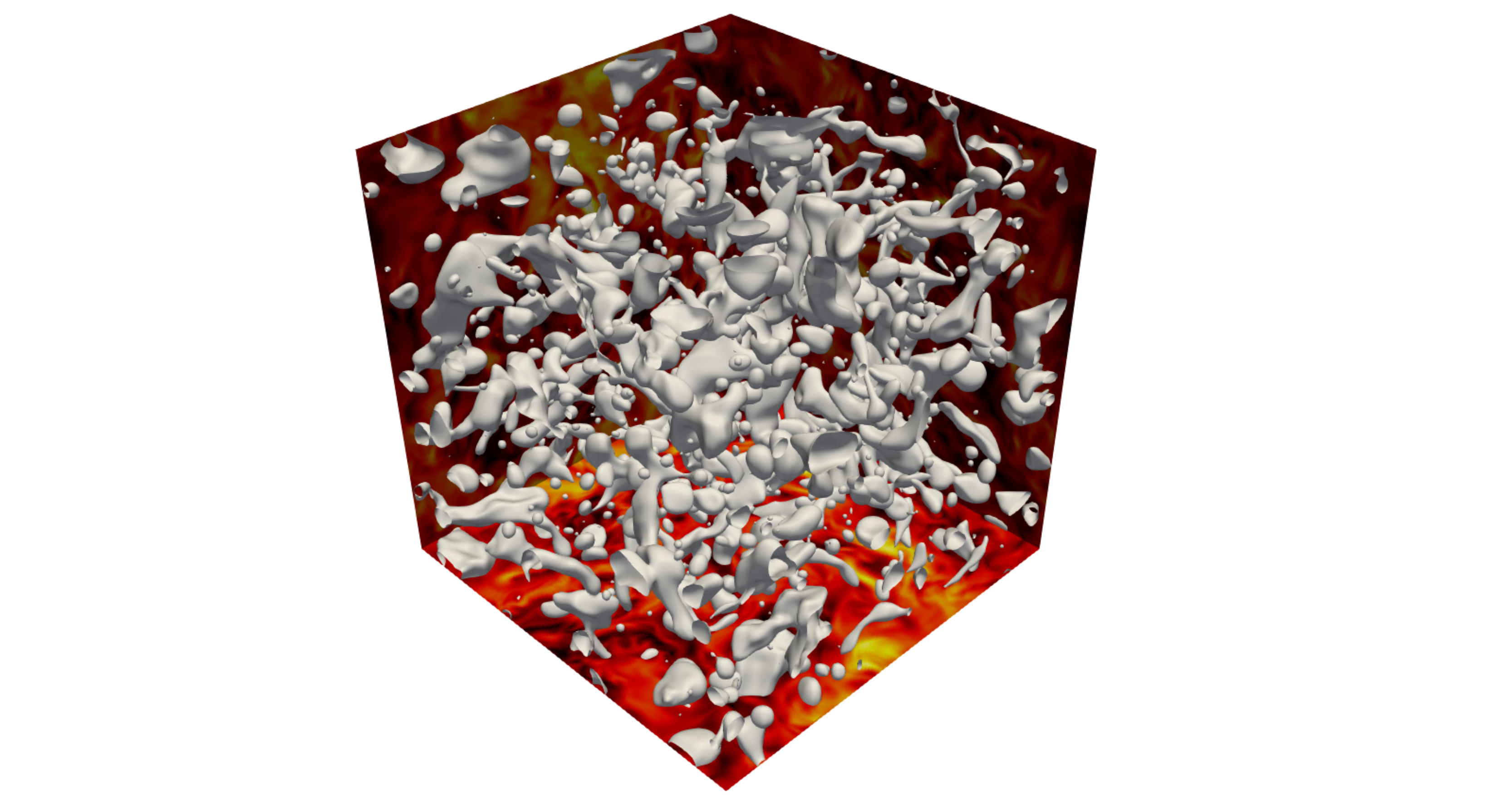}
    \caption{A visualization of the turbulent two-phase flow field showing a $\phi =0.5$ isocontour of the dispersed phase and the velocity magnitude on the faces.}
    \label{fig:paraview}
\end{figure}

To further assess the performance of the methods in more realistic turbulent circumstances, the three approaches in Section \ref{sec:approaches} were applied to a two-phase flow field with homogeneous isotropic turbulence from the work of \cite{jain2025stationary}. The domain consists of a cube and was initialized with a droplet at the center of the flow field. This was then subjected to a forced homogeneous isotropic turbulence and allowed to reach a dynamic stationary state. A snapshot of the flow field with the droplets in this stationary state is shown in Figure \ref{fig:paraview}. The parameters for this simulation were chosen to be a density of 1, a viscosity of $2.0334 \times 10^{-3}$ for both the dispersed and carrier phases, a void fraction of the dispersed phase of 0.0655, a Taylor-microscale Reynolds number of $Re_{\lambda} \approx 87$, and an integral-scale Weber number of $We_l = (2/3) \rho_c u_l^2 l_e/\sigma = 19.5$. Here, $\rho_c$ is the carrier phase density, $l_e$ is the integral length scale of the single-phase flow defined as $l_e = k_e^{3/2}/\epsilon$, where $k_e$ and $\epsilon$ are the kinetic energy and dissipation, respectively, and $u_l$ is the root-mean-square velocity, which is related to the kinetic energy through $k_e = 3/2u_l^2$. The computational setup consisted of a triply periodic domain of size $(2\pi)^3$ discretized using $N^3$ points with N = 256{, and the value of $\epsilon$ was chosen to be $0.51\Delta$}.

\begin{figure}[h!]
    \centering
    \includegraphics[width=0.9\linewidth]{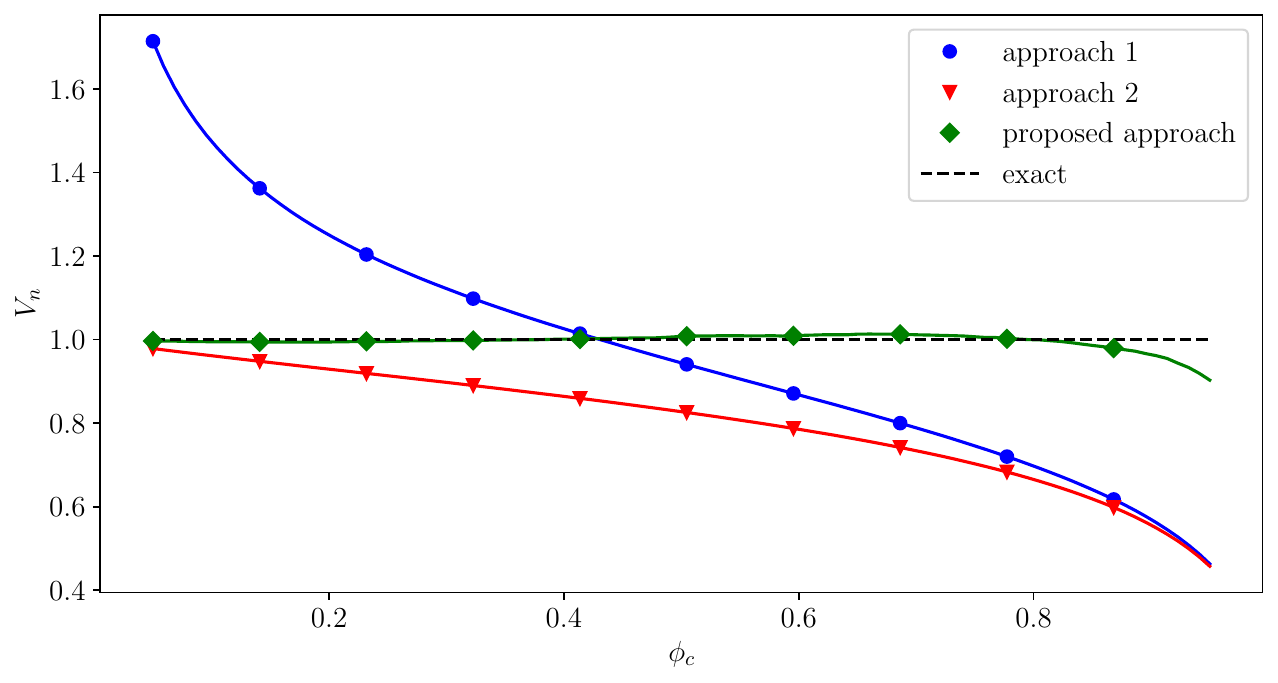}
    \caption{The calculated normalized total volume of all the droplets, $V_n$, as a function of volume fraction cut-off values for the turbulent flow field. The proposed approach outperforms the other approaches over all values of $\phi_c$.}
    \label{fig:turbresults}
\end{figure}

The three approaches to calculate the droplet volumes in Section \ref{sec:approaches} were applied to the snapshot of the turbulent two-phase flow in Figure \ref{fig:paraview}. Figure \ref{fig:turbresults} demonstrates the performance of each of the three approaches, by plotting $V_n$ over a range of $\phi_c$ values. {Here $V_n$ is the normalized total volume of all the droplets in the domain calculated using Equation \ref{eq:V_n}, where the normalization is done using the exact total volume of all the droplets computed using Equation \ref{eq:V_true}. This is a good estimation of the exact volume of all the droplets as long as there are no/minimal spurious droplets in the domain (see Appendix E).} Similar to the results in Section \ref{sec:results1}, approach 1 results in overprediction of the droplet volumes for lower values of $\phi_c$ and underprediction for higher values; approach 2 results in underprediction for all values of $\phi_c$; and the proposed approach accurately calculates the total droplet volume for all values of $\phi_c$, except a slight underprediction for very large values of $\phi_c$.

Table \ref{tab:resultstable} provides a more quantitative view of the performance of the approaches for a few specific values of $\phi_c$. The approach 1 is most accurate for $\phi_c$ of approximately 0.4, resulting in an error of $1.442\%$, but it diverges quickly when using other cutoff values. In contrast, the approach 2 is most accurate for smaller vales of $\phi_c$, and progressively gets worse as the chosen cutoff value increases. Note, albeit approach 2 is accurate enough for $\phi_c \rightarrow 0$, this is not a viable option because it results in artificial merging of the droplets as illustrated in Figure \ref{fig:fflimits2}(b). The proposed approach remains accurate for all values of $\phi_c$, maintaining an error of less than $1\%$.

\begin{table}[h]
    \centering
    \begin{tabular}{| c | c | c | c |}
      \hline
      \textbf{$\phi_c$} & \textbf{Approach 1} & \textbf{Approach 2} & \textbf{Proposed Approach} \\
      \hline
      0.2 & 20.38\% & -8.080\% & 0.4175\% \\
      \hline
      0.4 & 1.442\% & -14.05\% & 0.1262\% \\
      \hline
      0.5 & -5.922\% & -17.44\% & 0.8212\% \\
      \hline
      0.6 & -12.89\% & -21.27\% & 0.9219\% \\
      \hline
      0.8 & -28.01\% & -31.69\% & 0.1972\% \\
      \hline
    \end{tabular}
    \caption{Error, $E= V_n - V_{exact}$, in the calculated volume.}
    \label{tab:resultstable}
\end{table}

\subsubsection{Temporal evolution}

\begin{figure}[h!]
    \centering
    \includegraphics[width=\linewidth]{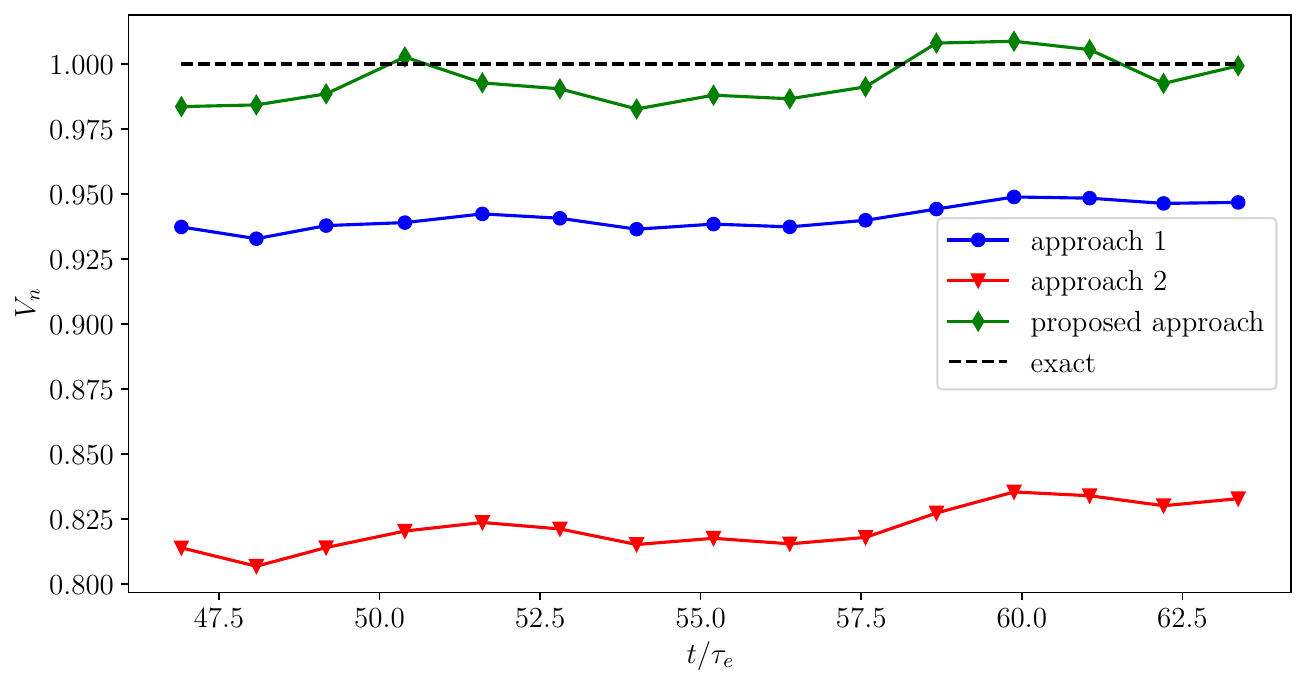}
    \caption{The calculated normalized total volume of all the droplets, $V_n$, over time in a turbulent flow for all three approaches for $\phi_c=0.5$. The proposed approach outperforms the other approaches for all time steps. Here, $\tau_e = {2k}/({3\epsilon})$ is the large eddy turnover time, where $k$ is the turbulent kinetic energy and $\epsilon$ is the turbulent dissipation.}
    \label{fig:tempanal}
\end{figure}

Next, the procedure in Section \ref{sec:results2} was repeated for multiple time steps of the same simulation to further assess the performance of the three approaches in Section \ref{sec:approaches}. 
Figure \ref{fig:tempanal} shows the total volume of the droplets as a function of time, calculated with the three approaches for $\phi_c=0.5$ (a typical value used in the flood-fill algorithms).
Since this is an incompressible two-phase flow with no phase change, the total volume of all the droplets should be conserved. The proposed approach recovers the exact volume to within $\approx2\%$ for all times. The variation in the volume captured by the method is likely due to the deformation of the droplets, affecting the errors borne from the spherical droplet assumption. The approaches 1 and 2 perform consistently over time, but do not recover the true volume at this value of $\phi_c$. 

\subsubsection{Droplet size distributions}

Next, to determine the acceptable range of $\phi_{c}$ for the calculation of flow statistics, an analysis on the number of droplets recovered by the approaches in Section \ref{sec:approaches} was done for the same two-phase turbulent flow field in Section \ref{sec:results2}. The number of droplets recovered varies significantly with the value of $\phi_c$ due to the artificial merging of drops (for low values of $\phi_c$) and missing some drops (for high values of $\phi_c$), with all three approaches recovering the same number of droplets. But the three approaches are not equally accurate in calculating droplet volumes for all values of $\phi_c$. The proposed approach is the most accurate approach for calculation of droplet volumes for a wide range of $\phi_c$ except for large values over $0.8$. Hence, a suitable range of $\phi_c$ needs to be determined that the proposed approach not only results in accurate droplet volumes but also can recover the number of droplets accurately. 

The droplets in the flow field were broken up into three size brackets. The smallest drops that are close to grid scale were considered small drops, which accounted for $0.01\%$ of the total volume of drops and hence are not counted. The medium-size drops represented those drops that contain upto about 1000 cells (average drop size in this simulation), and the large droplets represented everything larger than this value.

\begin{figure}[h!]
    \centering
    \includegraphics[width=\linewidth]{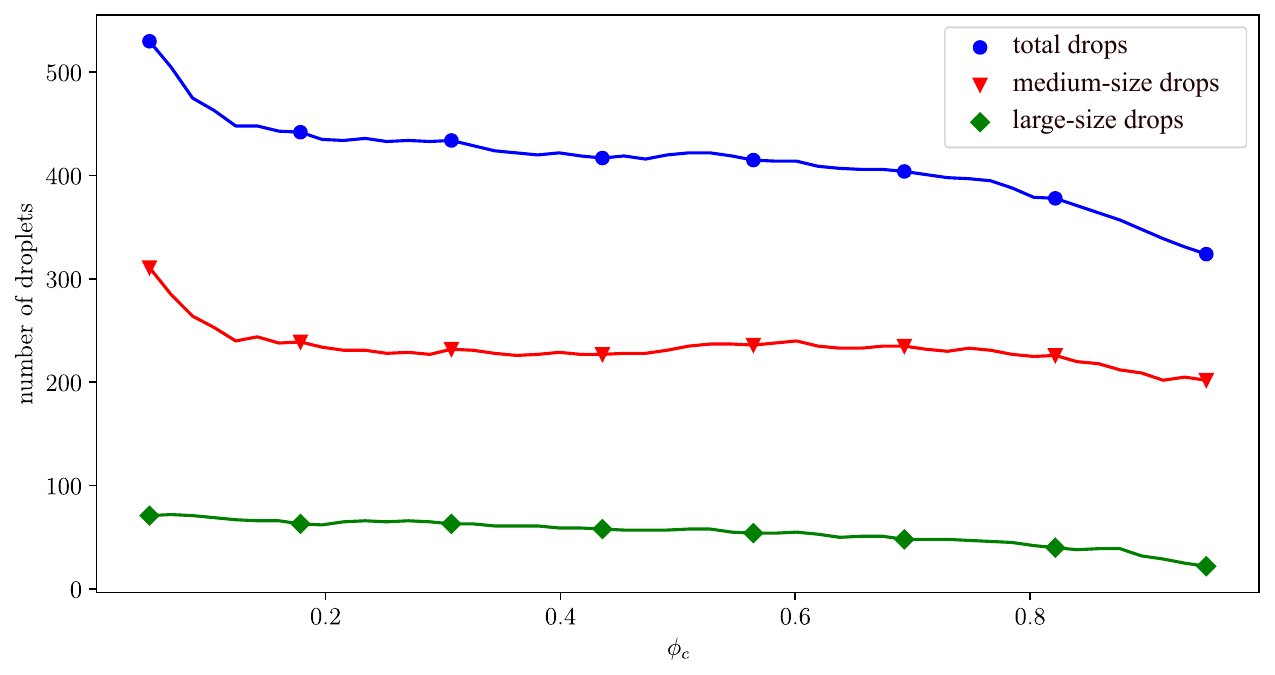}
    \caption{The number of large- and medium-sized droplets is shown as a function of the volume fraction cutoff value, along with their sum (total drops). The total number of droplets remains relatively constant for the range of $\phi_c$ from $0.2$ to $0.6$.}
    \label{fig:numdrops}
\end{figure}

Figure \ref{fig:numdrops} shows the number of droplets remains roughly invariant to changes in the $\phi_{c}$ value between the ranges of 0.2 to 0.6. There appears to be a drop-off in the number of droplets within the range of $\phi_c = 0$ to $0.2$, and another drop-off within the range of $\phi_c = 0.6$ to $1$. This lower drop-off range (0 - 0.2) is driven by the removal of really small droplets that were misclassified as medium drops due to the artificial merging that takes place for very small values of $\phi_c$. This is the reason the drop-off is only seen for the medium-sized drops. Hence, $\phi_c$ should not be picked in the range of $(0-0.2)$. The higher drop-off range $(0.6 - 1)$ could be explained by the removal of droplets that don't contain a section of volume fraction values that is high enough to classify them as a droplet. Hence, this range should also be avoided for the value of $\phi_c$. Overall, in order to avoid undercounting or overcounting the droplets within the flow while maintaining an accurate calculation of the total volume, the recommended range of $\phi_{c}$ values is $0.2$ to $0.6$. 

\begin{figure}[h!]
    \centering
    \includegraphics[width=\linewidth]{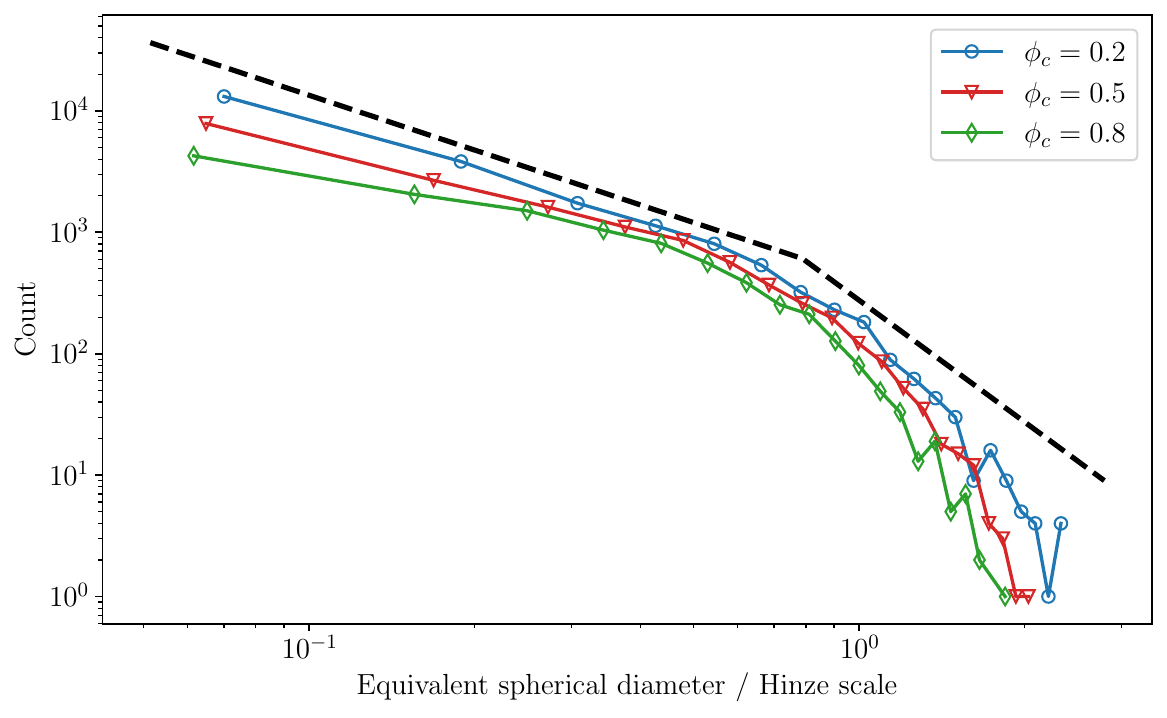}
    \caption{The distribution of droplet sizes, normalized by the Hinze scale, is plotted on a log-log scale for three different values of $\phi_c$ using approach 1. The dashed black lines represent the -10/3rd and -3/2th scaling laws.}
    \label{fig:distv1}
\end{figure}

\begin{figure}[h!]
    \centering
    \includegraphics[width=\linewidth]{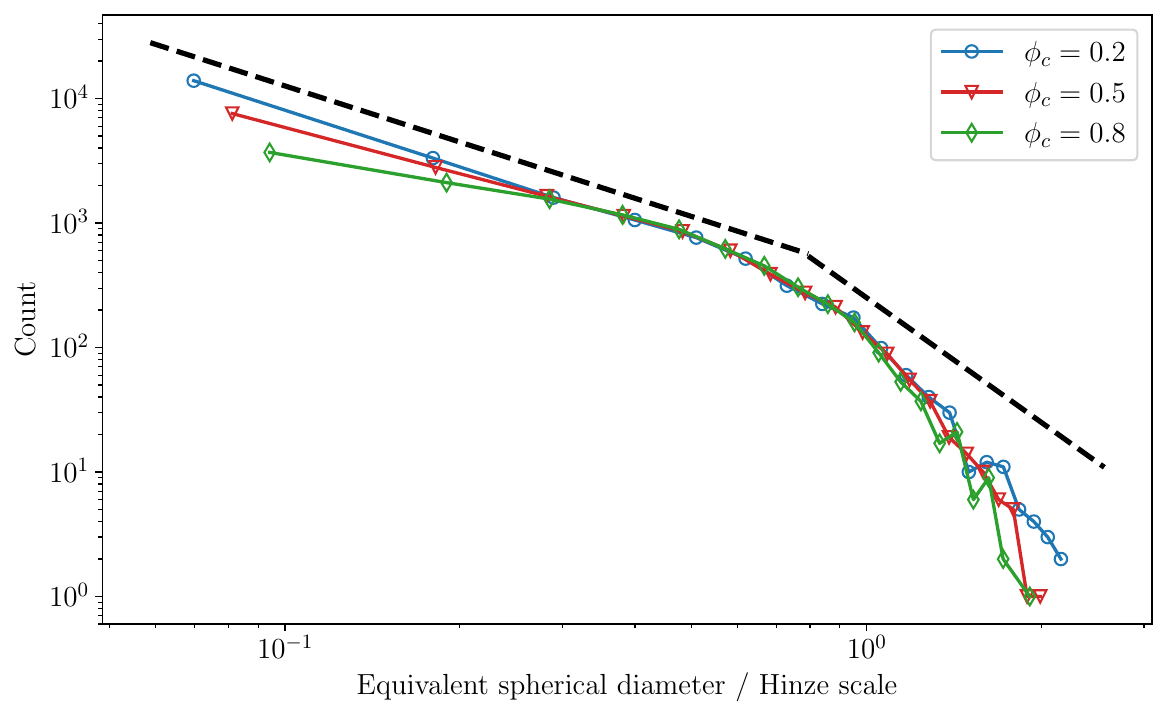}
    \caption{The distribution of droplet sizes, normalized by the Hinze scale, is plotted on a log-log scale for three different values of $\phi_c$ using the proposed approach. The dashed black lines represent the -10/3rd and -3/2th scaling laws.}
    \label{fig:distprop}
\end{figure}

{Additionally, to determine whether the changing cutoff value alters the droplet size distribution significantly in each of the approaches, the droplet size distributions for the same simulation case were plotted. Figures \ref{fig:distv1} and \ref{fig:distprop} plot a histogram of the equivalent spherical droplet diameter normalized by the Hinze scale of the flow. It can be seen that the distribution of the droplet sizes shifts up and down significantly based on the cutoff value for approach 1. However, the drop size distributions remain relatively unchanged for the proposed approach and also results in less scatter in the plots.}

\section{Concluding remarks\label{sec:conclusions}}

In this paper, a new approach for accurately calculating dispersed-phase properties (e.g., volume, number) in diffuse-interface two-phase flow simulations is outlined. Current flood-fill algorithms rely on an arbitrary volume-fraction cutoff value and often fail to accurately capture the diffuse interface region, resulting in significant volume errors. Using a simple analytical approximation of the truncated volume in the diffuse interface region, the proposed method corrects for these missing volumes.

The new approach was tested against two representative scenarios: a single spherical droplet and multiple droplets in a homogeneous isotropic turbulent flow field. In both cases, the proposed approach was shown to recover droplet volumes accurately across a wide range of volume-fraction cutoff values, offering significant improvements over the current approaches. This also permits one to choose a suitable volume fraction cutoff value that avoids artificial merging of nearby droplets while still recovering the accurate volume of drops, therefore, yielding a more accurate representation of the number of drops. Hence the proposed method enables more accurate processing of the two-phase flow solution fields. Additionally, the approach was tested in an evolving turbulent flow field, and it remained accurate over time.

Overall, the proposed method offers a simple and effective way to accurately calculate volumes in diffuse-interface two-phase flow simulations. This method can improve the calculation of droplet size distributions, total and individual dispersed-phase volumes, and droplet/bubble counts in two-phase flow simulations. Accurate calculation of these droplet and bubble properties is also important for the conversion of Eulerian-to-Lagrangian phase for Euler-Lagrange simulations of two-phase flows. 
Overall, this study aids in the design, optimization, and predictive modeling of complex two-phase systems in various engineering and environmental applications.

\section*{Acknowledgments}

P.~J.~N. acknowledges financial support from the President's Undergraduate Research Award (PURA) at Georgia Institute of Technology. S.~S.~J. acknowledges financial support from the George W. Woodruff School of Mechanical Engineering at Georgia Institute of Technology and the GWW-GTRI Connect Seed grant. S.~S.~J. also acknowledges the generous computing resources from the DOE's SummitPLUS award (TUR145, PI: Jain) and ALCC award (TUR147, PI: Jain). This research used resources of the Oak Ridge Leadership Computing Facility, which is a DOE Office of Science User Facility supported under Contract DE-AC05-00OR22725. 

\section*{Source code}
The source code of the programs/tools developed in this work will be available here on: 
\href{https://github.com/Flow-Physics-Computational-Science-Lab/diffuse-interface_post-process.git}{GitHub}

\appendix

\section{Ellipsoid vs spherical surface area approximation}

\begin{figure}[h!]
    \centering
    \includegraphics[width=\linewidth]{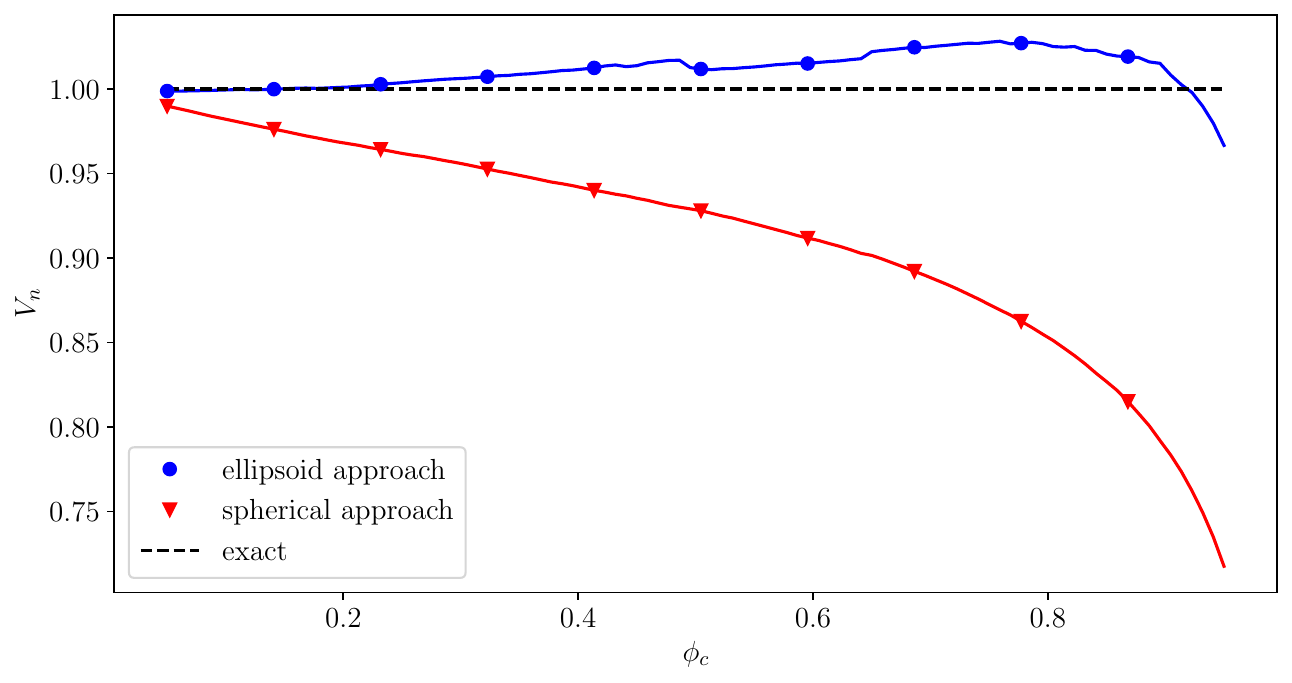}
    \caption{{The calculated normalized total volume of all the droplets, $V_n$, as a function of the volume fraction cut-off values from the turbulent flow field. The ellipsoid approach outperforms the spherical approach for all values of $\phi_c$}.}
    \label{fig:sphericalvellipsoid}
\end{figure}

{An alternate method of calculating the surface area of the droplets is using a spherical approximation rather than an ellipsoid approximation. To calculate the surface area of a droplet using this method, an equivalent spherical diameter is first calculated as
\begin{equation}
D_{eq,2} = 2\left(\frac{3V_{drop,2}}{4 \pi}\right)^\frac{1}{3},
\label{eq:dropletD}
\end{equation} 
where $V_{drop,2}$ is an estimation of the droplet volume computed using approach 2. The surface area is then computed using the expression
\begin{equation}
SA_{drop,2} = 4\pi \left(\frac{D_{eq,2}}{2}\right)^2.
\label{eq:dropletSA}
\end{equation} 
This surface area can be used in Equation \ref{eq:V_drop3} in place of the ellipsoid approximation to calculate the volume of the droplet. These two approaches are compared in Figure \ref{fig:sphericalvellipsoid}, where the ellipsoid method demonstrates a significant accuracy improvement over the spherical method for the droplets in the turbulence case in Section \ref{sec:results2}.
}

\section{2D version of the proposed method}

\begin{figure}[h!]
    \centering
    \includegraphics[width=\linewidth]{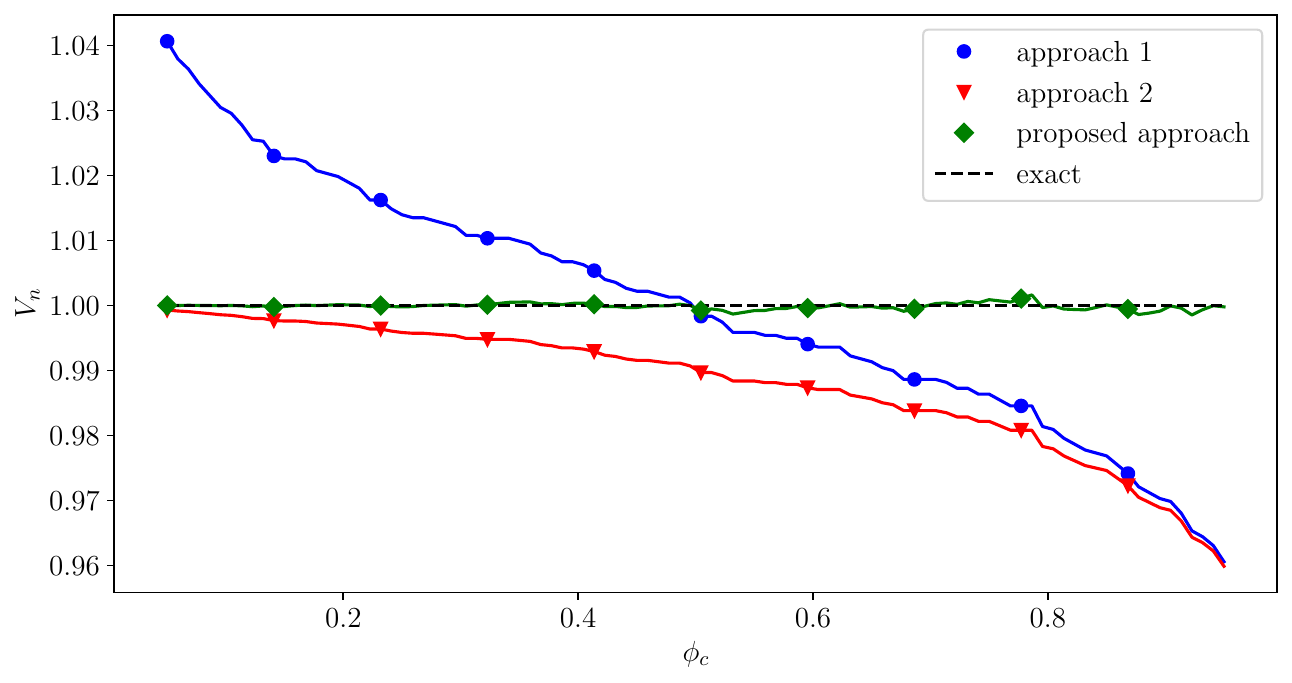}
    \caption{{The calculated normalized droplet volume, $V_n$, as a function of volume fraction cut-off values for the single 2D droplet. The proposed approach outperforms the other approaches for all values of $\phi_c$.}}
    \label{fig:2DAllApproaches}
\end{figure}

{To apply this methodology to a 2D simulation, slight modifications need to be made. To determine the perimeter (instead of the surface area) with which the interface thickness will be multiplied to, a simple arithmetic-geometric mean approximation can be used as
\begin{equation}
P_{drop,2} = 2\pi \sqrt{\frac{a^2 + b^2}{2}}.
\label{eq:dropletperi}
\end{equation} 
From there, the same methodology can be applied, using the same interface thickness term in Equation \ref{eq:integral} and by multiplying this to the perimeter to obtain an accurate 2D volume. This was applied and tested on a 2D version of the spherical droplet case from Section \ref{sec:results1}. As shown in Figure \ref{fig:2DAllApproaches}, the proposed approach outperforms the other methods for all values of $\phi_{c}$.
}

\section{Grid resolution study}
\label{sec:sample:appendix}
\begin{figure}[h!]
    \centering
    \includegraphics[width=\linewidth]{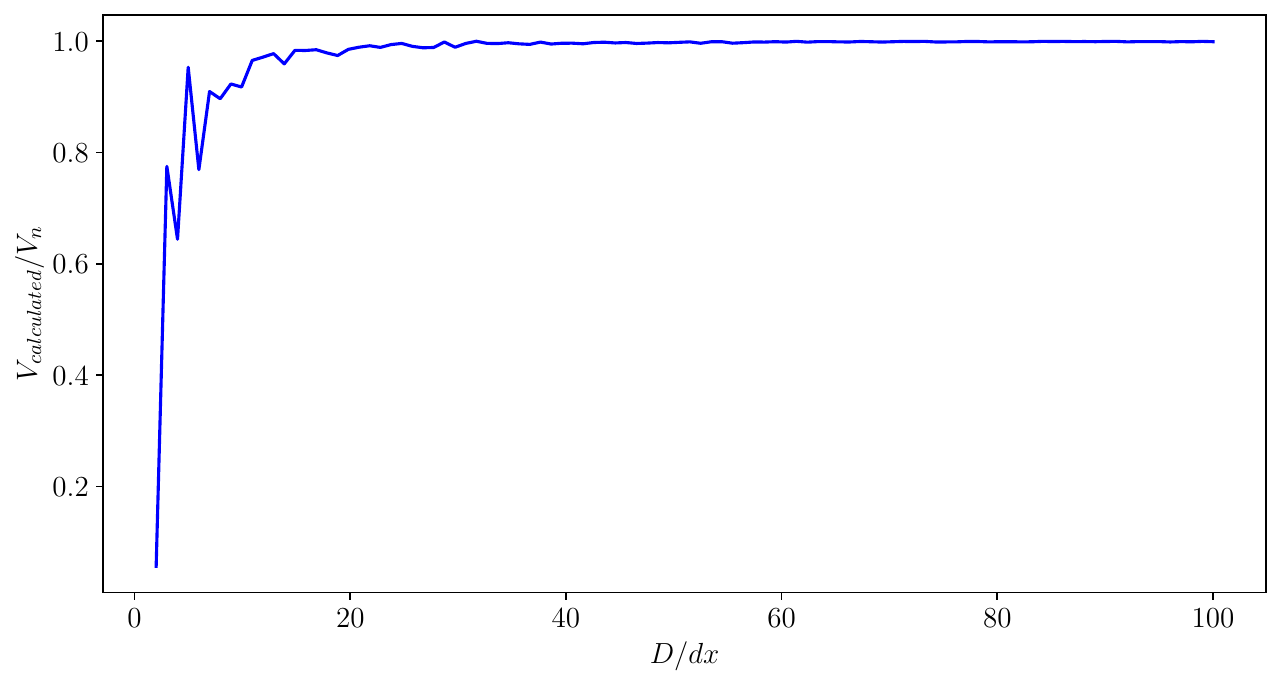}
    \caption{The accuracy $V_{calculated}/V_n$ is plotted against the ratio of droplet diameter $D$ to grid size $dx$ for the proposed approach.}
    \label{fig:gridsizing}
\end{figure}

The proposed approach requires that the droplets in the simulation are sufficiently resolved on the grid. If there are too few grid points per droplet, the accuracy of the method can drop. To determine this limit, a study on the relationship between droplet diameter and the accuracy of the model was done using the setup described in section \ref{sec:results1}. As can be seen in Figure \ref{fig:gridsizing}, there is a sharp drop off in the accuracy of the approach as the ratio of the diameter of the droplet $D$ to the grid size $dx$ drops below a value of approximately 20. It is therefore important to ensure that the droplets are resolved enough on the grid. But this is required not only for the accuracy of the proposed method, but more importantly to perform accurate numerical simulations of two-phase flows to accurately capture all the forces acting on the droplets. Hence, the proposed method is not imposing any new restrictions in terms of the grid resolution requirements.

\section{Interface thickness study}

\begin{figure}[h!]
    \centering
    \includegraphics[width=\linewidth]{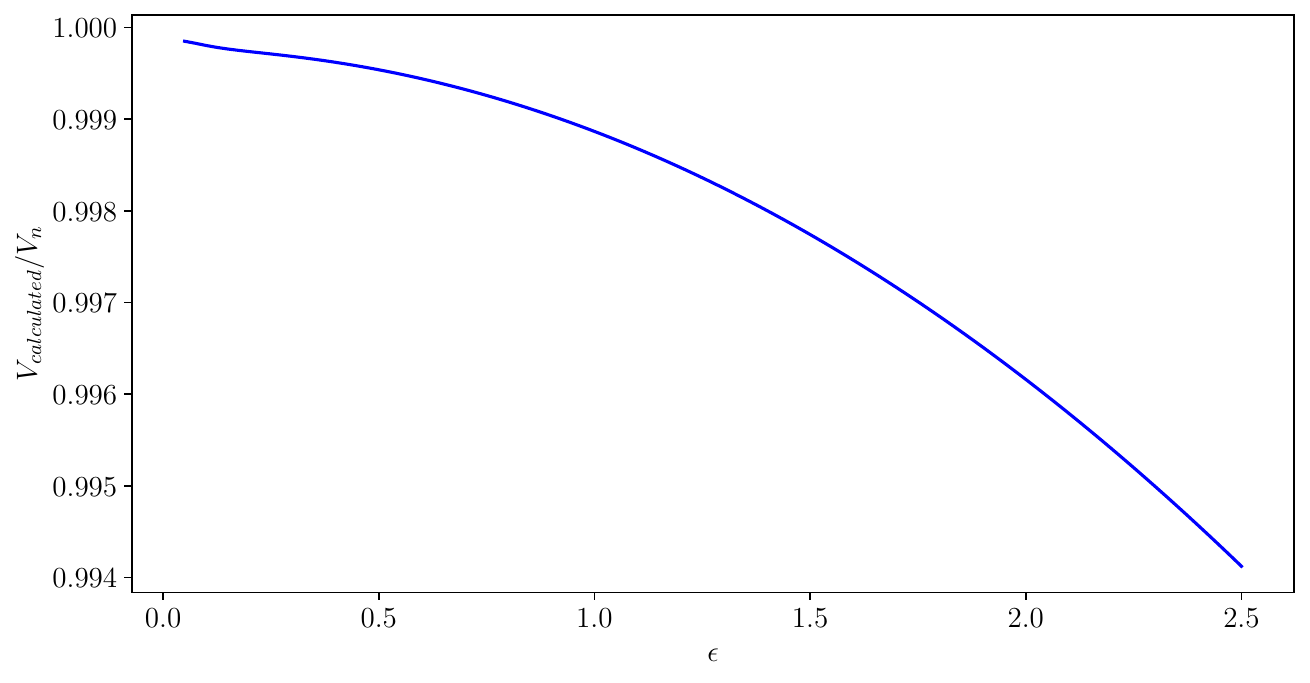}
    \caption{{The accuracy $V_{calculated}/V_n$ is plotted against the interface thickness $\epsilon$ for the proposed approach.}}
    \label{fig:interfacethickness}
\end{figure}

{To ensure the robustness of the proposed method across a wide range of interface thicknesses, the single droplet case in section \ref{sec:results1} was tested by varying the interface thickness, $\epsilon$. The volume of the droplet recovered is shown in Figure \ref{fig:interfacethickness} for various values of $\epsilon$, which shows that the method is more accurate for smaller values of $\epsilon$, but does not demonstrate a significant dropoff for larger values. Hence, the proposed approach is sufficiently robust to handle different interface thicknesses. However, the ACDI interface-capturing method was shown to be most accurate around a value of $\epsilon$ of $0.51\Delta$ in \citet{JAIN2022111529}; hence this is the value used throughout this study.}

\section{Spurious droplet characterization study}

To determine whether the $V_{exact}$ used as a true total volume in the flow does not include any significant contributions from spurious droplets, a methodology was developed to characterize the total volume of these droplets. A spurious droplet was defined as a droplet where the max volume fraction within the droplet was below a certain cutoff value, $\phi_{max}$. This methodology was applied to the simulation data from Section \ref{sec:results2} and plotted below. As shown in Figure \ref{fig:spurious}, the percentage of $V_{exact}$ that is in the spurious droplets increases as $\phi_{max}$ increases. However, this percentage remains below one percent up to about a $\phi_{max}$ of 0.9. This indicates that there is not significant volume in the spurious droplets and $V_{exact}$ is an accurate measure of the total volume.

\begin{figure}[h!]
    \centering
    \includegraphics[width=\linewidth]{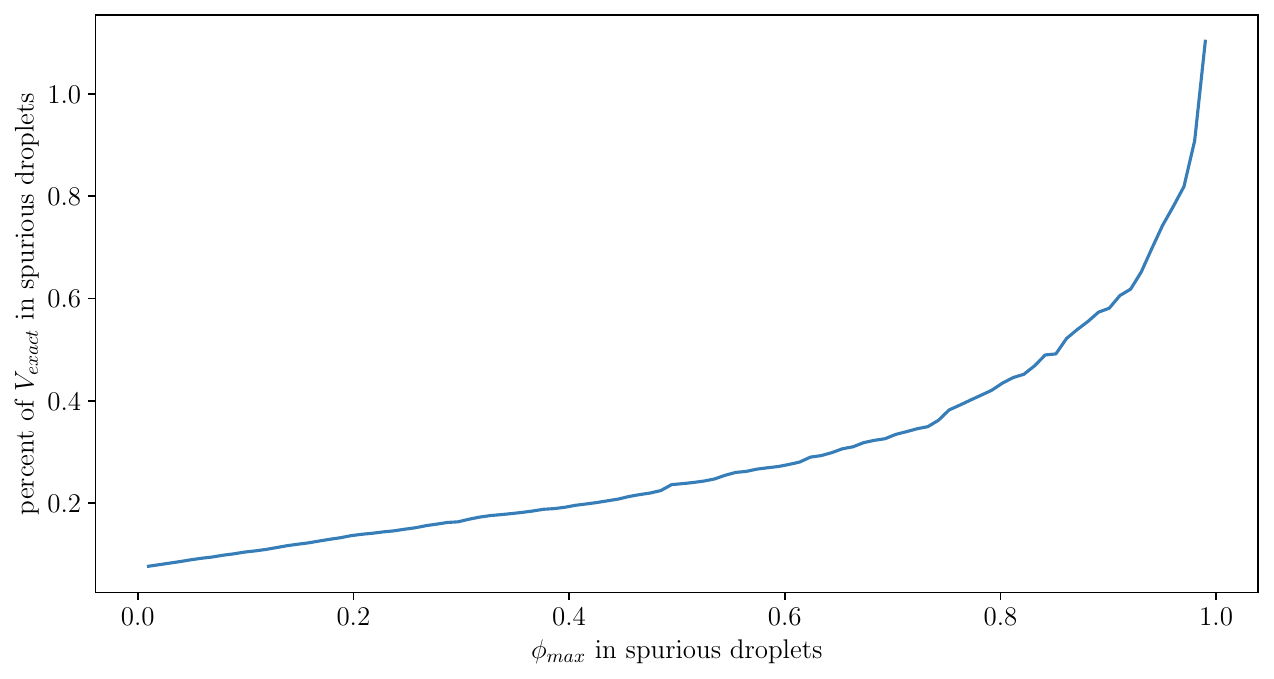}
    \caption{The percent of $V_{exact}$ that is within the spurious droplets is plotted against $\phi_{max}$ in spurious droplets.}
    \label{fig:spurious}
\end{figure}

 \bibliographystyle{model1-num-names} 
 \bibliography{cas-refs}





\end{document}